\documentclass[twocolumn, showpacs,preprintnumbers,amsmath,amssymb,prl]{revtex4}
\usepackage{amssymb}
\usepackage{amsfonts}
\usepackage{amsthm}
\usepackage{graphicx}% Include figure files
\usepackage{dcolumn}% Align table columns on decimal point
\usepackage{bm}% bold math
\usepackage{color}

\newcommand\mE{{\mathcal E}}
\newcommand\bs{{\boldsymbol s}}
\newcommand\bt{{\boldsymbol t}}
\newcommand\bc{{\boldsymbol c}}

\begin{document}

\title{$\frac{1}{2}+\frac{1}{2}>1$ for quantum error correction}

\author{Zhuo Wang$^{1,2}$\footnote{wangzhuo@iphy.ac.cn}}
\author{Sixia Yu$^2$\footnote{yusx@nus.edu.sg}}
\author{Heng Fan$^1$}
\author{C.H. Oh$^2$}
\affiliation{%
$^1$Institute of Physics, Chinese Academy of Sciences, Beijing
100190, China\\
$^2$Centre for Quantum Technologies, National University of
Singapore, 3
Science Drive 2, Singapore 117543}%
\date{\today}% It is always \today, today,
             %  but any date may be explicitly specified


\begin{abstract}
Quantum error correction assisted by entanglement helps to transmit the encoded qudits through quantum channels with some of them being noiseless. Here we consider a more realistic scheme for experiments what we called as {\it partial-noisy} quantum channels in which, instead of completely free of noise, only part of the qudit suffers no noise. In this scenario we show by an explicit example that two half-noisy channels work better than one noiseless channel, a queer phenomenon showing $\frac{1}{2}+\frac{1}{2}>1$. Our example also saturates a unified quantum Singleton bound, valid for the standard and entanglement-assisted codes as well. Furthermore, as illustrated by a mixed-alphabet code with half-noisy channels, a higher dimensional physical qudit can so improve the performance of a partial-noisy channel that it even outperforms a noiseless channel.

\end{abstract}
\maketitle

Quantum error correction \cite{shor,bennett,steane,knill,ekert,laflamme,gottesman} is of great importance to the quantum computational technologies since it provides the primary tool, the quantum error-correcting code (QECC), to fight against quantum noises. Entanglement is an essential resource that plays a key role in many quantum informational technologies \cite{ftc,qkd,ep}. The entanglement-assisted quantum error-correcting code (EAQECC) \cite{brun} combines those two ideas above by introducing some preexisting maximally-entangled pairs (ebits) between the encoder (Alice) and the decoder (Bob) before communication, where the qubits possessed by Bob suffer no errors. The advantage of EAQECC lies in the fact  that it can be easily constructed, namely from any classic linear codes that are not necessarily self-dual, and, more exhilaratingly, the entanglement not only behaves as an assistance but also improves encoding rate comparing to the corresponding standard QECC \cite{dong}.

A key step to implement the scheme of EAQECC  is to prepare ebits between encoder and decoder.  In order to share one ebit, one should prepare an entangled pair on Alice's side then send one qubit to Bob through a noiseless quantum channel. However, the noiseless quantum channel does not exist in practice. That is to say, EAQECC is only an ideal model with noiseless assumption which is difficult in experiment. This seems a fatal limitation of the scheme.

The theory of quantum error correction is based on the assumption that the error occurring on any physical qudit is unknown during the process of transmission. Thus a QECC should be designed to have the ability of detecting arbitrary errors on this qudit. However, there exist some specific dominant noisy processes in which the probability of some errors is much higher than that of others. Here we call this kind of processes {\it partial-noisy quantum channels}. For example, the well-known amplitude damping channel \cite{damping1,damping2}, which has been realized experimentally, is of this kind. A qubit transmitted though this channel only suffers two errors, i.e., $A_1=I+\sigma_z$ and $A_2=\sigma_x+i\sigma_y$ with $\sigma_x,\sigma_z,\sigma_y$ being Pauli operators, while there should be three errors in the general case. Leung {\it et al.} \cite{leung} first found a four-qubit code that can correct such one-bit errors, while the shortest standard QECC of distance 3 needs at least five qubits. The fact that codes dealing with a restricted set of errors can have a higher encoding rate than codes dealing with general errors inspires us to consider the partial-noisy channels instead of the noiseless channels for quantum error correction, which can also be more easily implemented experimentally.

From its definition the partial-noisy channels are quite general: any quantum channel that causes a particular dominant quantum noise process can be regarded as a partial-noisy channel. Encoding rate is the key criterion for the efficiency of any code construction. Then a question arises naturally: how efficient is the scheme of QECCs with partial-noisy channels compared with the scheme of EAQECCs? For this purpose we consider the following scenario.

Let us take qudits of $p^2$ levels as physical resources and any $p^2$-level qudit can be regarded as a composite two $p$-level subsystems. Denote the bit shift and phase shift operators of an $p$-level qudit by $X=\sum_{j\in\mathbb Z_p}|j+1\rangle\langle j|$ and $Z=\sum_{j\in\mathbb Z_p}\omega_p^j|j\rangle\langle j|$ with $\omega_p=e^{i\frac{2\pi}p}$ and $\mathbb Z_p$ being the ring of modular $p$, then group $\{\{X_1,Z_1\}\otimes\{X_2,Z_2\}\}$ forms an error basis of the $p^2$-level qudit. We introduce a special partial-noisy channel called as {\it half-noisy channel} that does not bring any error to one of the two subsystems when the $p^2$-level qudit is transmitted through, i.e., $\epsilon(X_1,Z_1)\gg\epsilon(X_2,Z_2)\approx0$ with $\epsilon$ being the error probability. Apparently, if both subsystems are well-protected during the process of transmission, i.e., $\epsilon(X_1,Z_1)\approx\epsilon(X_2,Z_2)\approx0$, it is reduced to a noiseless channel. We label a noiseless channel by \lq1\rq and a half-noisy channel by \lq$\frac{1}{2}$\rq.

As usual we denote by $((n,K,d;e))_q$ an EAQECC on $(n+e)$ qudits of $q$ levels, in which $e$ qudits suffer no errors, that has a $K$-dimensional encoding subspace detecting $(d-1)$-bit errors. Similarly we denote by $((n,K,d;e@\frac{1}{2}))_q$ a code of distance $d$ that employs $(n+e)$ $q$-level qudits among which $e$ qudits are transmitted though half-noisy channels to encode $K$ logical states. Then the comparison of the encoding rate should be made between two codes with parameters $((n-e,K,d;2e@\frac{1}{2}))_q$ and $((n,K',d;e))_q$ respectively with the largest possible $K$ and $K^\prime$, since these two scenarios use the same amount of resources ($n+e$ qudits) and forbid errors on the same number of subsystems ($2e$ subsystems). In other words, they have the same number of possible errors. If the ratio $K/K^\prime>1$ then the code with half-noisy channels is more efficient than the EAQECC.

Specifically we consider the following task. Alice is given five 9-level qudits and is asked to encode some quantum information in these physical resources that can correct any 1-bit errors and send to Bob via some quantum channels. There are two kinds of quantum channels for the system: i) one of the five channels is noiseless; ii) two of them are half-noisy channels. If Alice uses the first kind of channels, as we know, the best she can do is to use the optimal EAQECC $((4,9,3;1))_9$ \cite{dong} to encode 9 logical states. Instead if she uses the second kind of channels to send encoded information to Bob she can do better. As it turns out there exists a $((3,27,3;2@\frac{1}{2}))_9$ code, whose construction is given later, that encodes 27 logical states which is 2 times more than via the optimal EAQECC. That is to say, in the process of quantum error correction $\frac{1}{2}+\frac{1}{2}>1$ happens, which means two half-noisy channels can work better than one noiseless channel.

Now we turn to the construction of QECCs with half-noisy channels, in which we need a recently discovered method called {\it composite coding clique} \cite{wang} based on the theory of graph state \cite{graph1,graph2} and coding clique \cite{clique1,clique2}. In general, for an $((n,K,d;e@\frac{1}{2}))_{p^2}$ code, the system of $(n+e)$ $p^2$-level qudits can be divided into two subsystems each of which contains $(n+e)$ $p$-level qudits. Thus any half-noisy channel can be regarded as the composite of a noiseless channel for one subsystem and a general noisy channel for another. Let us label the system with a vertex set $V$ containing $(n+e)$ vertices. Denote by $P_1$ and $P_2$ two subsets of $V$ containing $e_1$ and $e_2$ vertices respectively that satisfy conditions $P_1\cap P_2=\varnothing$ and $e_1+e_2=e$. Then  $P_1$ and $P_2$ indicate the error-free qudits on the corresponding subsystems. A $\mathbb Z_p$-weighted graph $G_p=(V,\Gamma)$ defined on $V$ is a composed of a set of weighted edges specified by the adjacency matrix $\Gamma$ which is an $(n+e)\times(n+e)$ matrix with zero diagonal entries and the matrix element $\Gamma_{ab}\in\mathbb Z_p$ indicating the weight of the edge connecting vertices $a$ and $b$. The graph state on $G_p$ reads $|\Gamma\rangle=\frac1{\sqrt{p^n}}\sum_{\boldsymbol\mu\in \mathbb Z_p^V}\omega^{\frac12\boldsymbol\mu\cdot\Gamma\cdot\boldsymbol\mu}|\boldsymbol \mu\rangle$, where $|\boldsymbol\mu\rangle$ is the computational basis. Given two $\mathbb Z_p$-weighted graphs $G_1=(V,\Gamma_1)$ and $G_2=(V,\Gamma_2)$ for these two subsystems respectively, $\{Z_1^{\bc_1}|\Gamma_1\rangle\otimes Z_2^{\bc_2}|\Gamma_2\rangle\big|\bc_1,\bc_2\in\mathbb Z_p^V\}$ with $|\Gamma_1\rangle,|\Gamma_2\rangle$ being the corresponding graph states forms a basis of the system of $(n+e)$ $p^2$-level qudits. Define the {\it $(d,e_1,e_2)$-uncoverable set} as
\begin{eqnarray}
&& \mathbb D_d=\mathbb Z_p^V\otimes\mathbb Z_p^V-\{(\bt_1-\bs_1\cdot\Gamma_1)\otimes(\bt_2-\bs_2\cdot\Gamma_2)\big| \nonumber \\
&& (\widehat{\bs}_1\cup\widehat{\bt}_1)\cap P_1=\varnothing,(\widehat{\bs}_2\cup\widehat{\bt}_2)\cap P_2=\varnothing,\nonumber \\
&& 0<|\widehat{\bs}_1\cup\widehat{\bt}_1\cup\widehat{\bs}_2\cup\widehat{\bt}_2|<d\}\nonumber
\end{eqnarray}
and the {\it $(d,e_1,e_2)$-purity set} as
\begin{eqnarray}
&& \mathbb S_d=\{\bs_1\otimes\bs_2\Big|(\widehat{\bs}_1\cup\widehat{\bs_1\cdot\Gamma_1})\cap P_1=\varnothing, \nonumber \\
&& (\widehat{\bs}_2\cup\widehat{\bs_2\cdot\Gamma_2})\cap P_2=\varnothing,|\widehat{\bs}_1\cup\widehat{\bs_1\cdot\Gamma_1}\cup\widehat{\bs}_2\cup\widehat{\bs_2\cdot\Gamma_2}|<d\},\nonumber
\end{eqnarray}
where $\bs_1,\bs_2,\bt_1,\bt_2\in\mathbb Z_p^V$, $\widehat{\boldsymbol\mu}=\{a\in V|\mu_a\neq 0\}$ is the
support of vector $\boldsymbol\mu$ and $|C|$ indicates the number
of elements in $C\subseteq V$. We have

{\bf Theorem 1} An $((n,K,d;e@\frac{1}{2}))_{p^2}$ code with logical states
\begin{equation}
\{Z_1^{\bc_1}|\Gamma_1\rangle\otimes Z_2^{\bc_2}|\Gamma_2\rangle\Big|\bc_1\otimes\bc_2\in\mathbb C_d^K\}
\end{equation}
can be defined if there exists a composite coding clique $\mathbb{C}^K_d$ composed of $K$ vectors $\{\bc_1^i\otimes\bc_2^i|i=1,\cdots,K\}$ in $\mathbb Z_p^V\otimes\mathbb Z_p^V$ that satisfy:
\begin{itemize}
\item[(i)]  ${\bf 0}\in\mathbb{C}^{K}_d$;
\item[(ii)]  $\bs_1\cdot\bc_1+\bs_2\cdot\bc_2=0$ for all $\bs_1\otimes\bs_2\in\mathbb S_d$, $\bc_1\otimes\bc_2\in\mathbb C^K_d$;
\item[(iii)]  $(\bc^i_1-\bc^j_1)\otimes(\bc^i_2-\bc^j_2)\in\mathbb D_d$ for all $\bc^i_1\otimes\bc^i_2, \bc^j_1\otimes\bc^j_2\in\mathbb C^K_d$.
\end{itemize}

Using this construction we regard the 9-level system as two 5-qutrit subsystems. Given two $\mathbb Z_3$-weighted star graphs $\Gamma_5$ with all edges weighted one and the first and last vertices of the second graph being error free as shown in Fig.1(A), we can find a composite coding clique containing 27 elements that reads
\begin{equation}
\mathbb{C}^{27}_3=
\begin{pmatrix}
00000, \\
12220, \\
21110
\end{pmatrix}\otimes
\begin{pmatrix}
00000, &10000, &20000, \\
01112, &02220, &11110, \\
12220, &21110, &22222
\end{pmatrix},
\end{equation}
which defines a $((3,27,3;2@\frac{1}{2}))_9$ code. Complete details of this construction, along with backgrounds and rigorous proofs, can be found in the appendix.

\begin{figure}
\begin{center}
\includegraphics[width=2.8in]{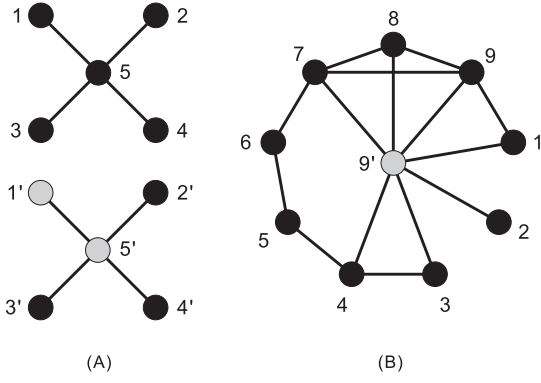}
\caption{(A) Graph for the $((3,27,3;2@\frac{1}{2}))_9$ code: two $\mathbb Z_3$-weighted star graphs are paired up, all edges are weighted one, and gray vertices label the error-free qutrits; (C) Graph of the $((8,2^5,3;1@\frac{1}{2}))_{2^84^1}$ code: vertices $9$ and $9'$ are paired up to be the 4-level qudit, and gray vertex labels the error-free qubit.}
\end{center}
\end{figure}

The optimality of the quantum codes is decided by the quantum bound. Considering that the $((4,9,3;1))_9$ code is optimal but our $((3,27,3;2@\frac{1}{2}))_9$ code can encode more logical states, QECCs with half-noisy channels should have a higher upper bound than the scheme of EAQECCs. Here we study the quantum Singleton bound \cite{knill,qSB} and try to give an unified expression for the following three schemes: QECCs with half-noisy channels, EAQECCs and standard QECCs. We denote by $((n,K,d;e,e'@\frac{1}{2}))_{p^2}$ a code of distance $d$ that employs $(n+e+e')$ $p^2$-level qudits among which $e$ and $e'$ qudits are transmitted through noiseless channels and half-noisy channels respectively to encode $K$ logical states. When $e=e'=0$, or $e=0$ and $e'\neq0$, or $e'=0$ and $e\neq0$, it is reduced to a standard QECC, a QECC with half-noisy channels and an EAQECC respectively. Then the calculation of the Singleton bound follows the Theorem in below.

{\bf Theorem 2} For an $((n,K,d;e,e'@\frac{1}{2}))_{p^2}$ code it holds
\begin{equation}\label{qSB}
K\leq
\left\{\begin{split}
&p^{n+2e+2e'-2(d-1)},2(d-1)>n\geq2(d-1)-e'\\
&p^{2n+2e+2e'-4(d-1)},n\geq2(d-1)
\end{split}\right.
\end{equation}

\begin{proof}
Partition $(n+e+e')$ qudits into 3 parts, i.e., part $A$, $B$ and $C$ containing $d-1$, $d-1$ and $n+e+e'-2(d-1)$ qudits respectively. All the $e$ noiseless qudits must be assigned into part $C$. Any half-noisy qudit in part $A$ or $B$ should be cut into two $p$-level halves and sacrifice the noiseless half to part $C$. These operations ensures that all possible errors occurring on part $A$ or $B$ labeled by $\{\varepsilon_\alpha\}$, $\{\varepsilon_\beta\}$ also form a basis of them. Denote the dimensions of the three parts by $K_A$, $K_B$ and $K_C$ respectively and introduce three reduced projectors which are $P_A=Tr_{BC}P$, $P_B=Tr_{AC}P$ and $P_{BC}=Tr_AP$ with $P$ being the projector of the encoding space, we have $P_A\propto\sum_{\alpha\subseteq A}Tr(P\varepsilon_\alpha)\varepsilon^\dagger_\alpha$. Thus $TrP_A^2=\sum_{\alpha\subseteq A}\frac{|Tr(\varepsilon_\alpha P)|^2}{K_A}$. Taking account of the error-correction condition that $P\varepsilon_i\varepsilon_j^\dagger P=\frac{1}{K}Tr(P\varepsilon_i\varepsilon_j^\dagger)P$, we have $\sum_{\alpha\subseteq A}\frac{|Tr(\varepsilon_\alpha P)|^2}{K_A}=\sum_{\alpha\subseteq A}\frac{Tr(\varepsilon_\alpha P\varepsilon^{\dagger}_\alpha P)}{K_A}=KTrP^2_{BC}\geq\frac{K}{K_C}TrP^2_B$, where the inequality is due to $Tr(P_{BC}-P_B/K_C)^2\geq0$. Thus $TrP_A^2\geq\frac{K}{K_C}TrP^2_B$. Similarly,  $TrP_B^2\geq\frac{K}{K_C}TrP^2_A$ holds for part $B$ as well. Then we can obtain an inequality that $K\leq K_C$. Different partitions may lead to different values of $K_C$ and the Singleton bound should be the smallest one that $K\leq K_{C,min}$.

Now we calculate the value of $K_{C,min}$. Firstly condition $n\geq2(d-1)-e'$ must be satisfied since both part $A$ and $B$ can not contain noiseless qudits. Secondly the less half-noisy qudits are assigned into part $A$ and $B$, the smaller the value of $K_C$ will be, since the noiseless halves of the half-noisy qudits in part $A$ and $B$ increase the dimensions of part $C$. When $n\geq2(d-1)$, all half-noisy qudits could be assigned into part $C$. Hence
\begin{equation}
K_{C,min}=(p^2)^{n+e+e'-2(d-1)}=p^{2n+2e+2e'-4(d-1)}.\nonumber
\end{equation}
When $2(d-1)>n\geq2(d-1)-e'$, part $A$ and $B$ contain at least $2(d-1)-n$ half-noisy qudits, which means part $C$ will be extended by $2(d-1)-n$ noiseless $p$-level halves at least. Thus we have
\begin{equation}
K_{C,min}=(p^2)^{n+e+e'-2(d-1)}\cdot p^{2(d-1)-n}=p^{n+2e+2e'-2(d-1)}.\nonumber
\end{equation}
\end{proof}

Now let us see how our scheme enhances the Singleton bound. According to the theory, if $e'=0$, then $n\geq2(d-1)$ no matter $e=0$ or not. That is to say, Singleton bound for both standard QECCs and EAQECCs is $K=p^{2n+2e-4(d-1)}$. If $e'\neq0$, then $n$ could be less than $2(d-1)$. In this case $K=p^{n+2e+2e'-2(d-1)}>p^{2n+2e+2e'-4(d-1)}$, which means QECCs with half-noisy channels is the best scheme among the three. Analyzing from the aspect of errors, our scheme and EAQECCs have a smaller number of possible errors than standard QECCs, thus they have a higher encoding rate. Moreover, our scheme reduces the number of short-bit errors comparing with the corresponding EAQECCs, hence it has a higher upper bound than EAQECCs although the total number of possible errors are the same. Now we look back on the example we present above. The Singleton bound for the $((4,K,3;1))_9$ code is $K=9$. However, since our scenario of two half-noisy channels has a smaller number of 1-bit and 2-bit errors, the Singleton bound is enhanced to $K=27$ which means our $((3,27,3;2@\frac{1}{2}))_9$ code is optimal as well.

We have shown that two half-noisy channels can work better than one noiseless channel in the process of quantum error correction. However, one may ask another question: is there any method that can give rise to one half-noisy channel working better than one noiseless channel? We find that the answer is positive as shown by the following example.

As we know, the optimal EAQECC of distance 3 with 8 qubits and one ebit is $((8,2^4,3;1))_2$ \cite{dong}. When the noiseless channel is replaced by a half-noisy channel, how to make the code encoding more logical states? Our scheme is to employ a 4-level qudit instead of a qubit to be transmitted through the half-noisy channel. Hence the code should be denoted by $((8,K,3;1@\frac{1}{2}))_{2^84^1}$. We can see that this is a QECC over mixed alphabets \cite{wang} as well since the physical resources for encoding have different number of energy levels. {\it Composite coding clique} is also the primary tool of constructing mixed-alphabet QECCs. Thus it should be involved to construct this $((8,K,3;1@\frac{1}{2}))_{2^84^1}$ code, as long as the corresponding positions of the uncoverable set and purity set are fixed to be error free. More details can be found in the appendix. Here we regard the 4-level qudit as a composite of two qubits one of which suffers no errors. Given a $\mathbb Z_2$-weighted graph $\Gamma_{10}$ on 10 vertices with the last vertex being error free as shown in Fig.1(B), we can find a composite coding clique generated by 5 generators that reads
\begin{eqnarray}
\mathbb{C}^{32}_3=&\{100110000\otimes0,010001010\otimes0,001010010\otimes0,\nonumber \\
&000100110\otimes0,000011001\otimes0\}.
\end{eqnarray}
Then the encoding space spanned by basis
\begin{equation}
\{Z^{\bc_1\otimes\bc_2}|\Gamma_{10}\rangle\Big|\bc_1\otimes\bc_2\in\mathbb C_3^{32}\}
\end{equation}
defines a more efficient $((8,2^5,3;1@\frac{1}{2}))_{2^84^1}$ code. That is to say, the performance of the half-noisy channel with a 4-level qudit becomes even better than the noiseless channel. The example tells us that a higher dimensional qudit can help to improve the performance of a partial-noisy channel. In other words, if we do not have enough partial-noisy channels in hand to transmit encoded informations, we should consider to employ some higher dimensional qudits to do  the encoding.

In this article we consider the situation of the partial-noisy channels for QECCs to overcome the fatal limitation of EAQECCs---non existence of the noiseless quantum channels. Surprisingly, we discover, via an example of $((3,27,3;2@\frac{1}{2}))_9$ code, that our scheme could be more efficient than the scheme of EAQECCs. More exhilaratingly, a unified Singleton bound is derived showing that the code we find is not unique. Our scheme truly has the higher upper bound than the other two schemes (EAQECCs and standard QECCs), which shows the great potential of our method of code construction.

There are many different kinds of partial-noisy channels depending on physical systems in laboratory,
and some of them may not be well described yet since that the exact implementation systems are not fixed.
The results of half-noisy channels which can be easily mapped to noiseless channels provide us a start point for more
complicated situations. This opens a new direction in studying QECCs in a more realistic background.
We believe that other kinds of partial-noisy channels can work better than the noiseless channels as well, some may even be more powerful than the half-noisy channels. Following the scheme in this paper, we need to identify the corresponding mappings
between different channels,  find methods to construct codes with other kinds of partial-noisy channels
and propose the corresponding quantum Singleton bounds.

The construction of the $((8,2^5,3;1@\frac{1}{2}))_{2^84^1}$ code shows another phenomenon that the cooperation of mixed-alphabet QECCs and partial-noisy channels could lead to better efficiency. Though it also brings higher complexity in constructions, the potential of the cooperation is worth being further explored.

This work is supported by 973 Program of China (Grant No. 2010CB922904), National Research Foundation and Ministry of Education of Singapore (Grant No. WBS: R-710-000-008-271) and NSF of China (Grant No. 11075227).

\newpage
\widetext

\appendix
\section{Backgrounds}
\subsection{Non-binary Graph States and Non-binary QECCs \cite{clique2}}
A $p$-level qudit is a particle with $p$ energy levels defined on the ring $\mathbb Z_p=\{0,1\ldots, p-1\}$. Denote the computational basis of a qudit by $\{|i\rangle|i\in \mathbb Z_p\}$, then the bit shift and phase shift operators are defined as
\begin{equation}
X=\sum_{l\in\mathbb Z_p}|l+1\rangle\langle l|,\quad
Z=\sum_{l\in\mathbb Z_p}\omega^{l}|l\rangle\langle
l|,\quad\left(\omega=e^{ i\frac {2\pi}p}\right).
\end{equation}
which satisfy $ZX=\omega XZ$ and $X^p=Z^p=I$. We can see that $X$ and $Z$ also form an error basis of the qudit. Here we introduce another important basis $\{|\theta_i\rangle|i\in \mathbb Z_p\}$ for a qudit which reads
\begin{equation}
|\theta_i\rangle=\frac1{\sqrt p}\sum_{l\in \mathbb
Z_p}\omega^{-li}|l\rangle.
\end{equation}
Then we have
\begin{equation}
X_i|\theta_i\rangle=\omega^i|\theta_i\rangle,\quad Z_i|\theta_i\rangle=|\theta_{i-1}\rangle,
\end{equation}
which means that $X$ becomes the phase shift operator and $Z$ becomes the bit shift operator for this basis.

Consider an $n$-qudit system. A $\mathbb Z_p$-weighted graph $G=(V,\Gamma)$ is composed of a set
$V$ of $n$ vertices and a set of weighted edges specified by the
{\it adjacency matrix} $\Gamma\in \mathbb Z_p^{n\times n} $, i.e., an
$n\times n$ matrix with zero diagonal entries and the matrix element
$\Gamma_{ab}\in\mathbb Z_p$ indicating the weight of the edge
connecting vertices $a$ and $b$. Denote by $\mathbb Z^V_p$ the set of all the vectors
$\boldsymbol\mu=(\mu_1\mu_2\cdots\mu_n)$ with $n$ components $\mu_a\in \mathbb Z_p$
$(a\in V)$, then the graph state of this weighted graph reads
\begin{equation}
|\Gamma\rangle=\frac1{\sqrt{p^n}}\sum_{\boldsymbol\mu\in \mathbb
Z_p^V}\omega^{\frac12\boldsymbol\mu\cdot \Gamma\cdot\boldsymbol\mu}|\boldsymbol\mu\rangle=\prod_{a,b\in V}\mathcal ({\mathcal
U}_{ab})^{\Gamma_{ab}}|\theta_0\rangle^{\otimes n},
\end{equation}
where $|\boldsymbol\mu\rangle$ is the computational bases of this system and $\mathcal U_{ab}$ is the non-binary controlled phase gate between qudits $a$ and $b$ that reads
\begin{equation}
\mathcal U_{ab}=\sum_{i,j\in\mathbb
Z_p}\omega^{ij}|i\rangle\langle i|_a\otimes |j\rangle\langle j|_b.
\end{equation}
Since $\mathcal U_{ab}$ satisfies
\begin{equation}
X_a^l\mathcal U_{ab}=Z_b^{-l}\mathcal U_{ab}X_a^l
\end{equation}
and $|\theta_0\rangle^V$ is the joint $+1$ eigenstate of $X_a$, we have
\begin{equation}
X_a|\Gamma\rangle=(\prod_{b\in V}Z_b^{-\Gamma_{ab}})|\Gamma\rangle.
\end{equation}
That is to say, any bit shift error acting on the graph state can be replaced by a phase shift error.
As we know, $\{\mE:=X^{\bs}Z^{\bt}| {\bs,\bt} \in \mathbb{Z}_{p}^V\}$ defines a nice error basis of the $n$-qudit system. For any error $X^{\bs}Z^{\bt}$ we have
\begin{equation}
X^{\bs}Z^{\bt}|\Gamma\rangle=\omega^{-\frac12\bs\cdot\Gamma\cdot\bs}Z^{\bt-\bs\cdot\Gamma}|\Gamma\rangle.
\end{equation}
Thus error $X^{\bs}Z^{\bt}$ can be replaced by $Z^{\bt-\bs\cdot\Gamma}$ up to some phase factors. When $\bt=\bs\cdot\Gamma$, $X^{\bs}Z^{\bs\cdot\Gamma}$ acting on $|\Gamma\rangle$ will be reduced to $I$. It means $|\Gamma\rangle$ is the joint $+1$ eigenstate of a stabilizer group
\begin{equation}
\{g^\bs:=X^{\bs}Z^{\bs\cdot\Gamma}\big|\bs\in\mathbb{Z}_{p}^V\}.
\end{equation}

We define a {\it d-uncoverable set} as
\begin{equation}
\mathbb D_d=\mathbb
Z_p^V-\{{\bt}-{\bs}\cdot\Gamma\Big|0<|\widehat{\bs}\cup\widehat{\bt}|<d\}
\end{equation}
and a {\it d-purity set} as
\begin{equation}
\mathbb S_d=\{{\bs}\in \mathbb
Z_p^V\Big||\widehat{\bs}\cup\widehat{\bs\cdot\Gamma}|<d\},
\end{equation}
where $\widehat{\boldsymbol\mu}=\{a\in V|\mu_a\neq 0\}$ is the
support of a vector $\boldsymbol\mu\in \mathbb Z_p^V$ and $|C|$ indicates the number
of elements in $C\subseteq V$.
A {\it coding clique} $\mathbb{C}^K_d$ of a given $\mathbb Z_p$-weighted graph
$G=(V,\Gamma)$ is a collection of $K$ different vectors in $\mathbb
Z_p^V$ that satisfy:
\begin{itemize}
\item[i)]  ${\bf 0} \in\mathbb{C}^{K}_d$;
\item[ii)]  ${\bs}\cdot{\bc}=0$ for all ${\bs}\in\mathbb S_d$ and
every ${\bc}\in \mathbb C^K_d$;
\item[iii)]  ${\bc}-{\bc}^\prime\in \mathbb D_d$ for all ${\bc}, {\bc}^\prime\in \mathbb C^K_d$.
\end{itemize}
Then the subspace spanned by
the basis $\{Z^{\bc}|\Gamma\rangle|{\bc}\in
\mathbb C^K_d\}$ forms an $((n, K, d))_p$ code.

\subsection{Composite Coding Clique and QECCs over Mixed Alphabets \cite{wang}}
A QECC over mixed alphabets is a code with physical particles for encoding having different number of energy levels. We denote a code over 2 alphabets by $((n,K,d))_{q^{n_1}p^{n_2}}$ which means the system has $n_1$ $q$-level qudits and $n_2$ $p$-level qudits with $n_1+n_2=n$. If $p$ is a divisor of $q$, i.e., $q=r\cdot p$, any $q$-level qudit ({\it quqit}) can be regarded as the composite of a $p$-level qudit ({\it qupit}) and an $r$-level qudit ({\it qurit}). Then the mixed-alphabet system can be regarded as the composite of an $n_1$-qurit subsystem and an $n$-qupit subsystem. Denote the bit shift and phase shift operators of a qupit and a qurit by $X_p,Z_p$ and $X_r,Z_r$ respectively, then $\{\{X_p,Z_p\}\otimes\{X_q,Z_q\}\}$ forms an error basis of a quqit. Given a $\mathbb Z_p$-weighted graph $G_p=(V,\Gamma_p)$ for the $p$-level subsystem and a $\mathbb Z_r$-weighted graph $G_r=(V_1,\Gamma_r)$ for the $r$-level subsystem where $V_1\subset V$ indicates the first $n_1$ vertices of vertex set $V$,
\begin{equation}
\{\mE_p\otimes\mE_r:=X_p^{\bs}Z_p^{\bt}\otimes X_r^{\bs'}Z_r^{\bt'}\big|{\bs,\bt} \in \mathbb{Z}_{p}^V,{\bs',\bt'} \in \mathbb{Z}_{r}^{V_1} \}
\end{equation}
forms a nice error basis of the mixed-alphabet system. Thus any less than $d$-bit error can be regarded as two errors occurring on these two subsystems respectively, i.e.,
\begin{equation}
|\mE|=|\mE_p\cup\mE_r|=|\widehat{\bs}\cup\widehat{\bt}\cup\widehat{\bs'}\cup\widehat{\bt'}|<d.
\end{equation}

We define the {\it d-uncoverable set} as
\begin{equation}
\mathbb D_d=\mathbb Z_p^V\otimes Z_r^{V_1}-\{({\bt}-{\bs}\cdot\Gamma_p)\otimes({\bt'}-{\bs'}\cdot\Gamma_r)\big|
0<|\widehat{\bs}\cup\widehat{\bt}\cup\widehat{\bs'}\cup\widehat{\bt'}|<d\}
\end{equation}
and the {\it d-purity set} as
\begin{equation}
\mathbb S_d=\{{\bs}\otimes{\bs'}\in \mathbb Z_p^V\otimes Z_r^{V_1}\Big|
|\widehat{\bs}\cup\widehat{\bs\cdot\Gamma_p}\cup\widehat{\bs'}\cup\widehat{\bs'\cdot\Gamma_r}|<d\}.
\end{equation}
A {\it composite coding clique} $\mathbb{C}^K_d$ is a collection of $K$ different vectors $\{\bc_i\otimes\bc'_i|i=1,\cdots,K\}$ in $\mathbb Z_p^V\otimes Z_r^{V_1}$ that satisfy:
\begin{itemize}
\item[(i)]  ${\bf 0} \in\mathbb{C}^{K}_d$;
\item[(ii)]  $\omega_p^{{\bs}\cdot{\bc}}\omega_r^{{\bs'}\cdot{\bc'}}=1$ for all ${\bs}\otimes{\bs'}\in\mathbb S_d$ and
 $\bc\otimes\bc'\in \mathbb C^K_d$;
\item[(iii)]  $(\bc_i-\bc_j)\otimes(\bc'_i\underline{\underline{}}-\bc'_j)\in \mathbb D_d$ for all $\bc_i\otimes\bc'_i, \bc_j\otimes\bc'_j\in \mathbb C^K_d$.
\end{itemize}
Then the subspace spanned by the basis
\begin{equation}
\{Z_p^{\bc}|\Gamma_p\rangle\otimes Z_r^{\bc'}|\Gamma_r\rangle\Big|\bc\otimes\bc'\in\mathbb C_d^K\}
\end{equation} forms an $((n,K,d))_{q^{n_1}p^{n_2}}$ mixed-alphabet code.

\section{proof of Theorem 1}
We need to prove that for any error that $0<|\mE|<d$, the logical states satisfy the Knill-Laflamme condition $\langle i|\mE|j\rangle=f(\mE)\delta_{ij}$. Firstly if $\mE$ is proportional to a stabilizer of the state $|\Gamma_1\rangle\otimes|\Gamma_2\rangle$, i.e., $\mE=f(\mE)\cdot g_1^{\bs_1}\otimes g_2^{\bs_2}$ with $f(\mE)$ being phase factor, since condition (ii) of the composite codeing clique ensures that $[X_1^{\bs_1}Z_1^{\bs_1\cdot\Gamma_1}, Z_1^{\bc_1}]=0$ and $[X_2^{\bs_2}Z_2^{\bs_2\cdot\Gamma_2}, Z_2^{\bc_2}]=0$ for any $\bc_1\otimes\bc_2\in\mathbb C^K_d$, we have
\begin{equation}
\begin{split}
&\langle i|\mE|j\rangle\\
=&f(\mE)\langle\Gamma_1|Z_1^{-\bc_1^i}\cdot g_1^{\bs_1}\cdot Z_1^{\bc_1^j}|\Gamma_1\rangle
\langle\Gamma_2|Z_2^{-\bc_2^i}\cdot g_2^{\bs_2}\cdot Z_2^{\bc_2^j}|\Gamma_2\rangle\\
=&f(\mE)\langle\Gamma_1|Z_1^{\bc_1^j-\bc_1^i}\cdot g_1^{\bs_1}|\Gamma_1\rangle
\langle\Gamma_2|Z_2^{\bc_2^j-\bc_2^i}\cdot g_2^{\bs_2}|\Gamma_2\rangle\\
=&f(\mE)\langle\Gamma_1|Z_1^{\bc_1^j-\bc_1^i}|\Gamma_1\rangle\langle\Gamma_2|Z_2^{\bc_2^j-\bc_2^i}|\Gamma_2\rangle\\
=&f(\mE)\delta_{ij}.
\end{split}
\end{equation}
Secondly if $\mE$ is not a stabilizer of state $|\Gamma_1\rangle\otimes|\Gamma_2\rangle$, then
\begin{equation}
\begin{split}
&\langle i|\mE|j\rangle\\
=&f(\mE)\langle\Gamma_1|Z_1^{\bc_1^j-\bc_1^i}\cdot X_1^{\bs_1}Z_1^{\bt_1}|\Gamma_1\rangle\langle\Gamma_2|Z_2^{\bc_2^j-\bc_2^i}\cdot X_2^{\bs_2}Z_2^{\bt_2}|\Gamma_2\rangle\\
=&f(\mE)\langle\Gamma_1|Z_1^{\bc_1^j-\bc_1^i}Z_1^{\bt_1-\bs_1\cdot\Gamma_1}|\Gamma_1\rangle\langle\Gamma_2|Z_2^{\bc_2^j-\bc_2^i}Z_2^{\bt_2-\bs_2\cdot\Gamma_2}|\Gamma_2\rangle\\
=&0
\end{split},
\end{equation}
where the third equality is due to the fact that condition (iii) of the composite coding clique makes at least one of $\bc_1^j-\bc_1^i+\bt_1-\bs_1\cdot\Gamma_1\neq0$ and $\bc_2^j-\bc_2^i+\bt_2-\bs_2\cdot\Gamma_2\neq0$ hold. Finally the Knill-Laflamme condition is satisfied. Thus the encoding space defines an $((n,K,d;e@\frac{1}{2}))_{p^2}$ code.


\begin{thebibliography}{99}

\bibitem{shor}P. W. Shor, {\it Phys. Rev. A} {\bf 52}, 2493 (1995).

\bibitem{bennett}C. H. Bennett, D. P. DiVincenco, J. A. Smolin and W. K. Wootters, {\it Phys. Rev. A} {\bf 54},
3824 (1996).

\bibitem{laflamme}R. Laflamme, C. Miquel, J.-P. Paz, and W. H. Zurek, {\it Phys. Rev. Lett.} {\bf 77}, 198 (1996).

\bibitem{steane}A. M. Steane, {\it Phys. Rev. Lett.} {\bf 77}, 793 (1996).

\bibitem{ekert}A. Ekert and C. Macchiavello, {\it Phys. Rev. Lett.} {\bf 77}, 2585 (1996).

\bibitem{gottesman}D. Gottesman, Caltech Ph.D Thesis (1997), eprint: quant-ph/9705052.

\bibitem{knill}E. Knill and R. Laflamme, {\it Phys. Rev. A} {\bf 55}, 900 (1997).

\bibitem{ftc} E. Knill, R. Laflamme, and W. H. Zurek, {\it Science} {\bf 279}, 342 (1998);
D. Gottesman, {\it Phys. Rev. A} {\bf 57}, 127 (1998).

\bibitem{qkd} C.H. Bennett and  G. Brassard, {\it Proceedings of IEEE International Conference on Computers,
Systems, and Signal Processing}, 175 (1984); A. K. Ekert, {\it Phys. Rev. Lett.} {\bf 67}, 661 (1991).

\bibitem{ep} S. Glancy, E. Knill, and H. M. Vasconcelos, {\it Phys. Rev. A} {\bf 74}, 032319 (2006).

\bibitem{brun} T. Brun, I. Devetak, and M.-H. Hsieh, {\it Science} {\bf 314}, 436 (2006).

\bibitem{dong}Y. Dong, X. Deng, M. Jiang, Q. Chen, and S. Yu, {\it Phys. Rev. A} {\bf 79},
042342 (2009).

\bibitem{damping1}W. H. Louisell, {\it Quantum Statistical Properties of Radiation} (Wiley, New York, 1973).

\bibitem{damping2}C. W. Gardiner, {\it Quantum Noise} (Springer-Verlag, New York, 1991).

\bibitem{leung}D. W. Leung, M. A. Nielsen, I. L. Chuang, and Y. Yamamoto, {\it Phys. Rev. A} {\bf 56},
2567 (1997).

\bibitem{wang}Z. Wang, S. Yu, H. Fan, and C. H. Oh, e-print arXiv:1205.4253.

\bibitem{graph1}M. Hein, J. Eisert, and H. J. Briegel, {\it Phys. Rev. A} {\bf 69},
062311 (2004).

\bibitem{graph2}D. Schlingemann, and R. F. Werner, {\it Phys. Rev. A} {\bf 65},
012308 (2001).

\bibitem{clique1} S. Yu, Q. Chen, and C. H. Oh, e-print arXiv:0709.1780.

\bibitem{clique2} D. Hu, W. Tang, M. Zhao, Q. Chen, S. Yu, and C.H. Oh, {\it Phys. Rev. A} {\bf 78}, 012306 (2008).

\bibitem{qSB}S. Yu, C. H. Lai, and C. H. Oh, e-print arXive:1005.4758.

\end{thebibliography}
\end{document}